\documentclass[preprintnumbers,amsmath,amssymb]{revtex4}
\usepackage{graphicx}
\usepackage{amsmath}
\usepackage{amsfonts}
\usepackage{amssymb}
\usepackage{bm}
\newcommand{\ve}{\mathbf}

\begin{document}

\title{Analysis of Dynamic Axial-Symmetric Shells}
\author{Burak Himmeto\={g}lu}
\affiliation{Physics Department, Bilkent University, Ankara,
Turkey.}

\begin{abstract}
The aim of this work is to analyze the dynamical behavior of
relativistic infinite axial-symmetric shells with flat interior
and a radiation filled curved exterior spacetimes. It will be
proven, by the use of conservation equations of Israel, that the
given configuration does not let expansion or collapse of the
shell which was proposed before, but rather the shell stays at
constant radius. The case of null-collapse will also be considered
in this work and it will be shown that the shell collapses to zero
radius, and moreover, if cylindrical flatness is imposed a
boundary layer is obtained still contrary to previous works.
\end{abstract}

\maketitle

\section{Introduction}

Dynamic relativistic shells were studied by many authors for many
different applications. Israel, who first formulated the tools for
analyzing shell models ~\cite{I} also worked on the collapse of
spherically symmetric shells ~\cite{I2} that separate an interior
Minkowski spacetime with an exterior Schwarzschild spacetime and
the possible violations of casuality in such systems. Later
Barrab\'{e}s and Israel ~\cite{Ba}-~\cite{BaI} developed the null
shell formalism and considered collapse of spherically symmetric
shells along with Poisson ~\cite{Po}. Later, Olea and
Cris\'{o}stomo treated the dynamics of shell models by the use of
the Hamiltonian formalism and considered collapse of spherically
symmetric shells ~\cite{CO}. Recently, Jezierski ~\cite{JJ}
studied the geometry of crossing null shells which is another
important application. Similar dynamic shell models prove to be
very important in wormhole constructions and such models are
extensively studied by Visser and he collected and summarized
these in his book ~\cite{Viss}. These wormhole models were
constructed by the so called \emph{cut and paste technique} where
two copies of the same spacetime are identified on a hypersurface
which is the boundary of two regions cut from the spacetimes. This
junction hypersurface in the spherically symmetric case that forms
the throat of the wormhole was given a radial degree of freedom in
order to minimize the use of stress-energy which violates the weak
energy condition. Recently, Eiroa and Simeone analyzed the case of
cylindrical cut and paste wormholes ~\cite{ES} which is the direct
generalization of Visser's approach to axial-symmetric spacetimes.
Axial-symmetric dynamic shells were also studied by Ger\u{s}l,
Klep\'{a}\u{c} and Horsk\'{y} who considered a charged cylindrical
shell that separates an interior Bonnor-Melvin Universe and an
exterior Datta-Raychaudhuri spacetime ~\cite{GKH}. Other examples
of dynamic axial-symmetric shells were studied by Pereira and Wang
~\cite{PW} and then later by Seriu ~\cite{MS}, who considered an
axial-symmetric dynamic shell which separates an interior flat
spacetime and an exterior radiation-filled curved spacetime.

Our consideration is based on the same configuration analyzed by
Pereira, Wang ~\cite{PW} and Seriu ~\cite{MS} before. Since the
most recent one is the work by Seriu, let us recall this work
~\cite{MS}. First he considered a general axial-symmetric
spacetime and calculated the elements of the second fundamental
form and the surface stress-energy tensor from the Israel
equations. Then he considered the more special case analyzed
before by Pereira and Wang and obtained dynamical equations
concerning the motion of the shell from the components of the
surface stress-energy tensor. By analyzing these dynamical
equations he concluded with the result that the configuration at
hand does not let the collapse of the shell even in the limits
where the radial velocity of the shell approaches to the speed of
light. (i.e. the null limit) Finally he compared his work with the
previous one by Pereira and Wang.

In our work we also use the same configuration used previously by
Pereira-Wang ~\cite{PW} and Seriu ~\cite{MS}, but the method we
use to obtain the dynamical equations concerning the motion of the
shell is totally different. After we calculate the elements of the
second fundamental form for both the interior and exterior
spacetimes, we directly use the conservation equations of Israel
and we prove that in the given configuration the shell stays at
constant radius and does not move at all. We also compare our
results with those obtained by Seriu in two ways. We first use the
null-shell formalism ~\cite{BaI}-~\cite{Po}-~\cite{Ba} directly
and show that the results obtained are different from those
obtained by him. Secondly, we showed that his results are
inconsistent with the conservation equations.

Our work possesses the use of conservation equations for
determining the dynamical behavior of thin shells which has a
non-vacuum exterior spacetime for the first time. Previously, only
the vacuum conservation equations were used by Israel
~\cite{I}-~\cite{I2} for considering dynamic spherical shells, so
this work is an example where the full conservation equations are
used.

In section II, we start directly by considering the null shell
limit in the configuration given by Seriu, Pereira and Wang
~\cite{MS}-~\cite{PW}. We first calculate the elements of the
transverse curvature in both spacetimes and then apply the
Barrab\'{e}s-Israel null shell formalism and obtain the surface
stress-energy tensor. We show that in this case the shell
collapses to zero radius, which is contrary to the recent work by
Seriu ~\cite{MS}. Moreover, when cylindrical flatness is imposed,
the shell turns into a boundary layer which is also contrary to
the above mentioned work.

In section III, we analyze the timelike case by using the Israel
formalism directly. We first calculate the elements of the second
fundamental form in both spacetimes and then use them in the
conservation equations which proves that the shell must stay
stationary at some constant radius. In addition, we compare our
results with the previous ones and show that the previous ones do
not reach the conclusions they made, when conservation equations
are used.

In section IV, we give a conclusion of our results.

Through out this work, we use the sign conventions of MTW
~\cite{MTW} except the sign of the second fundamental form and
transverse curvature is Israel's ~\cite{I}-~\cite{Po}. We use
$\Sigma$ to denote timelike hypersurfaces and $\Xi$ to denote null
hypersurfaces. $\ve{k}$ denotes the normal vector to $\Xi$,
$\ve{N}$ denotes the the transverse vector, $\sigma_{AB}$ denotes
the intrinsic metric on $\Xi$, $y^{a}=(\lambda,\theta^{C})$
denotes the intrinsic coordinates on $\Xi$, $C_{ab}$ denotes the
transverse curvature and finally upper case Latin indices runs
from 2 to 3.

\section{Cylindrical Shells: The Lightlike Limit}

We will work on the two spacetimes $\mathcal{M}^{-}$ and
$\mathcal{M}^{+}$ given in ~\cite{MS} separated by the
hypersurface $\Sigma$ in the lightlike limit. Let us give the
metrics corresponding to $\mathcal{M}^{-}$ and $\mathcal{M}^{+}$:
\begin{eqnarray}
ds_{+}^{2} &=&
e^{2\gamma(t_{+}-r)}(-dt^{2}_{+}+dr^{2})+dz^{2}+r^{2}d\phi^{2}
\label{-manif} \\ ds_{-}^{2} &=&
-dt_{-}^{2}+dr^{2}+dz^{2}+r^{2}d\phi^{2} \label{+manif}
\end{eqnarray}
In our case, the shell moves only radially so with respect to
$\mathcal{M}^{\pm}$, the null shell $\Xi$ is defined by the
equation $r=\rho(t_{\pm})$. Since the shell moves at the speed of
light, the radial component of 4-velocity satisfies,
\begin{equation}\nonumber
\frac{d\rho}{d\eta}=-1
\end{equation}
where $\eta$ is an affine parameter for the null geodesic of our
interest. The minus sign reflects the fact that the shell is
collapsing. Therefore, in $\mathcal{M}^{-}$, the collapsing shell
satisfies the equation,
\begin{equation}\nonumber
t_{-}+\rho=\vartheta_{-}=\textrm{constant}
\end{equation}
On the other hand, in $\mathcal{M}^{+}$ the equation for the null
geodesic that describes the collapsing shell $\Xi$ can be
calculated by solving the \emph{Euler-Lagrange Equations}.
Defining,
\begin{equation}\label{2K}
2K\equiv{e^{2\gamma}t_{+}^{\circ2}-e^{2\gamma}\rho^{\circ2}}=0
\end{equation}
where $q^{\circ}$ denotes $\frac{dq}{d\eta}$. The Euler-Lagrange
equations are given by,
\begin{equation}\label{EL}
\frac{\partial{K}}{\partial{q}}-\frac{d}{d\eta}\frac{\partial{K}}{\partial{q^{\circ}}}=0
\end{equation}
where $q$ represents the coordinates $t_{+},r,z,\phi$. Solving
this equation for $q=t_{+}$,
\begin{equation}\nonumber
\dot{\gamma}(\underbrace{t_{+}^{\circ2}-\rho^{\circ2}}_{0})-\frac{d}{d\eta}(e^{2\gamma}t_{+}^{\circ})=0
\end{equation}
where $\dot{\gamma}$ represents $\frac{d\gamma}{dt_{+}}$. Then we
are left only with the term $e^{2\gamma}t_{+}^{\circ}$ which is
equal to a constant $C$ because its derivative with respect to
$\eta$ is zero. Then we have,
\begin{equation}\label{tdot}
t_{+}^{\circ}=Ce^{-2\gamma}
\end{equation}
Using the above equality in (\ref{2K}) we get,
\begin{equation}\label{rdot}
\rho^{\circ}={\pm}Ce^{-2\gamma} \rightarrow
\rho^{\circ}=-Ce^{-2\gamma} \quad (\textrm{since we are
considering collapse})
\end{equation}
Then noting the following relation,
\begin{equation}
\frac{\rho^{\circ}_{+}}{t^{\circ}}=\frac{d\rho}{dt_{+}}=-1
\end{equation}
where in the last step we used (\ref{tdot}) and (\ref{rdot}).
Solving this trivial equation gives the equation of $\Xi$ in
$\mathcal{M}^{+}$ as,
\begin{equation}\nonumber
t_{+}+\rho=\vartheta_{+}=\textrm{constant}
\end{equation}
similar to its counterpart in $\mathcal{M}^{-}$. If we assume that
as $r{\rightarrow}{\infty}$,  $\mathcal{M}^{+}$ approaches to the
flat Minkowski metric (i.e as $r{\rightarrow}{\infty}$,
$\gamma{\rightarrow}0$ so $e^{2\gamma}{\rightarrow}1$), the
constant C should be chosen as 1. This case we refer to as the
cylindrically flat exterior spacetime.

Now let us use the \emph{Null-Shell Formalism} for the current
case:

\subsection{Transverse Curvature in {$\mathcal{M}^{-}$}}
Since $\frac{d\rho}{d\eta}=-1$ let us choose the parameter
$\lambda$ that describes the shell's null character as
$\lambda=\eta$, then intrinsic metric on the hypersurface $\Xi$ is
given by,
\begin{equation}\label{-metric}
ds_{\Xi}^{2}=dz^{2}+\lambda^{2}d\phi^{2}
\end{equation}
so $\Xi$ is described by the equation,
\begin{eqnarray}
t_{-} &=& \lambda + \vartheta_{-} \nonumber \\
r &=& -\lambda \nonumber \\
z &=& z \nonumber \\
\phi &=& \phi \nonumber
\end{eqnarray}
where in the above equations, right hand sides correspond to
intrinsic coordinates on $\Xi$, while the left hand sides show
their relation to coordinates in $\mathcal{M}^{-}$, so they
together define $\Xi$. Therefore, the tangent-normal vector of
$\Xi$ is given by,
\begin{equation}\label{k-}
k_{-}^{\alpha}=\frac{\partial{x_{-}^{\alpha}}}{\partial{\lambda}}=(1,-1,0,0)
\end{equation}
by using the equations we derived above that describes $\Xi$. Now
following ~\cite{Po}, we have to find the transverse vector
$\ve{N}$ that points out of $\Xi$ and different from $\ve{k}$
which must satisfy the following equations,
\begin{itemize}
\item[1)] $N_{\alpha}^{-}k^{\alpha}_{-}=-1$ \item[2)]
$N_{\alpha}^{-}N_{-}^{\alpha}=0$ \item[3)] $
N_{\alpha}^{-}e_{A}^{\alpha}=0$
\end{itemize}
Then the $3^{rd}$ condition indicates a form
$N_{\alpha}=(x,y,0,0)$ where $x$ and $y$ are to be found by
solving conditions 1 and 2 by using (\ref{k-}), which gives,
\begin{equation}\label{trans-}
N_{\alpha}=(-\frac{1}{2},\frac{1}{2},0,0)
\end{equation}
Then using the definition of \emph{Transverse Curvature},
\begin{eqnarray}
C_{ab} &=& -\ve{N}\cdot{}^{(4)}\nabla_{a}\ve{e_{b}} =
-N_{\alpha}e^{\beta}_{a}{}^{(4)}\nabla_{\beta}e_{b}^{\alpha}
\nonumber \\ &=&
-N_{\alpha}(\frac{\partial^{2}{x^{\alpha}}}{\partial{y^{a}}\partial{y^{b}}}+\Gamma_{\beta\gamma}^{\alpha}\frac{\partial{x^{\beta}}}{\partial{y^{a}}}\frac{\partial{x^{\gamma}}}{\partial{y^{b}}})
\label{openextrinsic}
\end{eqnarray}
Therefore, the only non-vanishing component the transverse
curvature becomes,
\begin{equation}\nonumber
C_{\phi\phi}^{-}=-N_{r}\Gamma_{\phi\phi}^{r}=\frac{r}{2}=\frac{r^{2}}{2r}=\frac{\sigma_{\phi\phi}}{2r}
\end{equation}
Therefore,
\begin{equation}\label{phiphi-}
C_{\phi\phi}^{-}\mid_{\Xi}=\frac{\sigma_{\phi\phi}}{2\rho}
\end{equation}

\subsection{Transverse Curvature in ${\mathcal{M}^{+}}$} The
treatment in $\mathcal{M}^{+}$ is totally similar. We again set
$\lambda=\eta$. The intrinsic metric on $\Xi$ is given in
$\mathcal{M}^{+}$ as,
\begin{equation}\nonumber
ds_{\Xi}^{2}=dz^{2}+\lambda^{2}d\phi^{2}
\end{equation}
since $e^{2\gamma}(-dt_{+}^{2}+d\rho^{2})=0$, same with
(\ref{-metric}) as it should be since the continuity of the metric
must be satisfied. Therefore $\Xi$ is described by the set of
equations,
\begin{eqnarray}
t_{+} &=& -\rho(\lambda)+\vartheta_{+} \nonumber \\
r &=& \rho(\lambda) \nonumber \\
z &=& z \nonumber \\
\phi &=& \phi \nonumber
\end{eqnarray}
Then the tangent-normal vector $k^{\alpha}$ is given as,
\begin{equation}\nonumber
k^{\alpha}=\frac{\partial{x_{+}^{\alpha}}}{\partial{\lambda}}=(-\frac{d\rho}{d\lambda},\frac{d\rho}{d\lambda},0,0)
\end{equation}
Since we calculated $\frac{d\rho}{d\lambda}$ in (\ref{rdot}) as
$\rho^{\circ}=-Ce^{-2\lambda}$, $k^{\alpha}$ becomes,
\begin{equation}\label{k+}
k_{+}^{\alpha}=Ce^{-2\gamma}(1,-1,0,0)
\end{equation}
Again we have to find the transverse vector $\ve{N}$ by the before
given properties,
\begin{equation}\nonumber
N_{\alpha}^{+}N_{+}^{\alpha}=0 \quad
N_{\alpha}^{+}k_{+}^{\alpha}=-1 \quad
N_{\alpha}^{+}e_{A}^{\alpha}=0
\end{equation}
which can also in view of $3^{rd}$ equation can be found as
putting $N_{\alpha}^{+}=(x,y,0,0)$ into the first two equations
which yields,
\begin{equation}\label{trans+}
N_{\alpha}^{+}=\frac{e^{2\gamma}}{2C}(-1,1,0,0)
\end{equation}
Then the non-vanishing components of the transverse curvature are
found also by (\ref{openextrinsic}).
\begin{eqnarray}
C_{\lambda\lambda}^{+} &=&
-N_{t}^{+}\{\frac{\partial^{2}{t_{+}}}{\partial{\lambda^{2}}}+\Gamma_{tt}^{t}(\frac{\partial{t_{+}}}{\partial{\lambda}})^{2}+\Gamma_{rr}^{t}(\frac{\partial{r}}{\partial{\lambda}})^{2}+2\Gamma_{rt}^{t}(\frac{\partial{r}}{\partial{\lambda}})(\frac{\partial{t_{+}}}{\partial{\lambda}})\}
\nonumber \\ &+&
-N_{r}^{+}\{\frac{\partial^{2}{r}}{\partial{\lambda^{2}}}+\Gamma_{tt}^{r}(\frac{\partial{t_{+}}}{\partial{\lambda}})^{2}+\Gamma_{rr}^{r}(\frac{\partial{r}}{\partial{\lambda}})^{2}+2\Gamma_{rt}^{r}(\frac{\partial{r}}{\partial{\lambda}})(\frac{\partial{t_{+}}}{\partial{\lambda}})\}
\label{C-ll1}
\end{eqnarray}
Now using the fact that $\gamma\equiv{\gamma(t_{+}-r)}$, we have
$\dot{\gamma}=-\gamma^{\prime}$ where prime denotes
differentiation with respect to $r$. So the Christoffel symbols
become for (\ref{+manif}),
\begin{equation}\label{C2}
\Gamma_{rr}^{r}=-\Gamma_{tt}^{t}=\Gamma_{tr}^{t}=-\Gamma_{rr}^{t}=\Gamma_{tt}^{r}=-\Gamma
_{tr}^{r}=\gamma^{\prime}
\end{equation}
Now since we have
$\frac{\partial{t_{+}}}{\partial{\lambda}}=-\frac{\partial{r}}{\partial{\lambda}}=Ce^{-2\lambda}$,
we get,
\begin{eqnarray}
\frac{\partial^{2}{t_{+}}}{\partial{\lambda^{2}}} &=&
-2\frac{d\gamma}{d\lambda}Ce^{-2\gamma}=-2C(\frac{d{\gamma}}{dr}\frac{dr}{d{\lambda}}+\frac{d{\gamma}}{dt_{+}}\frac{dt_{+}}{d{\lambda}})e^{-2\gamma}
\nonumber \\ &=&
4C^{2}\gamma^{\prime}e^{-4\gamma}=-\frac{\partial^{2}{r}}{\partial{\lambda^{2}}}
\label{2nder}
\end{eqnarray}
Thus putting (\ref{2nder}) and (\ref{C2}) into (\ref{C-ll1}) one
gets simply,
\begin{equation}\label{C-ll}
C_{\lambda\lambda}^{+}=0
\end{equation}
The only non-vanishing component of $C_{ab}^{+}$ then becomes,
\begin{equation}\nonumber
C_{\phi\phi}^{+}=-N_{r}^{+}\Gamma_{\phi\phi}^{r}=\frac{r^{2}}{2rC}
\end{equation}
Thus,
\begin{equation}\label{phiphi+}
C_{\phi\phi}^{+}\mid_{\Xi}=\frac{\sigma_{\phi\phi}}{2\rho{C}}
\end{equation}
Then we have,
\begin{eqnarray}
\mu &=& -\frac{1}{8\pi}\sigma^{AB}[C_{AB}] \nonumber \\
j^{A} &=& -\frac{1}{8\pi}\sigma^{AB}[C_{\lambda{B}}] \nonumber \\
p &=& -\frac{1}{8\pi}[C_{\lambda\lambda}]
\end{eqnarray}
with the surface stress-energy tensor given as,
\begin{equation}\label{surface}
S^{\alpha\beta}={\mu}k^{\alpha}k^{\beta}+j^{A}(k^{\alpha}e_{A}^{\beta}+k^{\beta}e_{A}^{\alpha})+p\sigma^{AB}e_{A}^{\alpha}e_{B}^{\beta}
\end{equation}
Since
\begin{equation}\nonumber
[C_{\lambda\lambda}]=0, \quad [C_{\lambda{B}}]=0, \quad
[C_{\phi\phi}]=\frac{\sigma_{\phi\phi}}{2\rho}(\frac{1-C}{C})
\end{equation}
We have,
\begin{equation}\label{mu}
\mu=\frac{C-1}{16{\pi}C\rho}
\end{equation}
Therefore,
\begin{equation}\nonumber
S^{\alpha\beta}=\frac{C-1}{16{\pi}C\rho}k^{\alpha}k^{\beta}
\end{equation}
From the above equations, we see that $C^{\pm}_{\lambda\lambda}=0$
which proves that the parameter chosen is affine both in
$\mathcal{M}^{+}$ and in $\mathcal{M}^{-}$ by ~\cite{Po}. So we
get a null shell with nonzero energy density, which collapses with
the speed of light to zero radius. On the other hand, if
cylindrical flatness (i.e. as $r{\rightarrow}{\infty}$,
$\gamma{\rightarrow}0$) is imposed as a special case (C=1), then
we get $\mu=\j^{A}=p=0$ so the hypersurface $\Xi$ becomes a
boundary layer rather than a thin shell.

\section{Collapse of Cylindrical Shell in Timelike Case}

In this section, we will work again on the spacetime described by
the metrics (\ref{-manif}) and(\ref{+manif}) but this time the
hypersurface connecting $\mathcal{M}^{-}$ and $\mathcal{M}^{+}$
will be timelike (i.e its normal vector $\ve{n}$ is spacelike), so
will be denoted by $\Sigma$. Again $\Sigma$ is defined by the
equation $\Sigma: \ r=\rho_{\pm}(t_{pm})$, which gives the induced
metric on $\Sigma$ as,
\begin{equation}\label{induced}
d_{\Sigma}^{2}=-d\tau^{2}+dz^{2}+r^{2}d\phi^{2}
\end{equation}
where
\begin{eqnarray}\label{propertime}
d\tau^{2} &=& e^{2\gamma}(1-\dot{\rho_{+}}^{2})dt_{+}^{2}
\nonumber \\ &=& (1-\dot{\rho_{-}}^{2})dt_{-}^{2} \nonumber
\end{eqnarray}
again dot represents derivative with respect to $t_{+}$ or
$t_{-}$. Defining the quantity
$\Delta^{-1}\equiv{\frac{dt_{+}}{dt_{-}}}$ and noting that
$\dot{\rho_{-}}=\Delta^{-1}\dot{\rho_{+}}$, we find from the above
equation,
\begin{eqnarray}
e^{2\gamma}(1-\dot{\rho_{+}}^{2})\Delta^{-2} &=&
1-\Delta^{-2}\dot{\rho_{+}}^{2} \nonumber \\ \Delta^{-1} &=&
[(1-e^{2\gamma})\dot{\rho_{+}}^{2}+e^{2\gamma}]^{-1/2}
\label{delta}
\end{eqnarray}
by imposing the continuity of the metric at $\Sigma$. In the
future, we will need to write $\ddot{\rho_{-}}$ in terms of the
quantities of $\mathcal{M}^{+}$. For this purpose, we know derive
some useful equations as follows(remember that
$\gamma^{\prime}=-\dot{\gamma}$) :

From equation (\ref{delta})
\begin{eqnarray}
\frac{d}{dt_{+}}\Delta &=&
\frac{1}{2}\Delta^{-1}\{2(1-e^{2\gamma})\dot{\rho_{+}}\ddot{\rho_{+}}+2\gamma^{\prime}e^{2\gamma}\dot{\rho_{+}}^{2}-2\gamma^{\prime}e^{2\gamma}\}
\nonumber \\ &=&
\frac{1}{\Delta}\dot{\rho_{+}}\ddot{\rho_{+}}-\frac{e^{2\gamma}}{\Delta}[\dot{\rho_{+}}\ddot{\rho_{+}}+\gamma^{\prime}(1-\dot{\rho_{+}}^{2})]
\label{ddelta}
\end{eqnarray}
Therefore,
\begin{equation}\nonumber
\ddot{\rho_{-}} = \frac{d}{dt_{-}}(\rho_{-})
\end{equation}
Using (\ref{ddelta}) above we get
\begin{eqnarray}
\frac{d}{dt_{-}}(\rho_{-}) &=&
\Delta^{-1}\frac{d}{dt_{+}}(\Delta^{-1}\dot{\rho_{+}}) \nonumber
\\ &=&
\frac{\ddot{\rho_{+}}}{\Delta^{2}}-\frac{\ddot{\rho_{+}}}{\Delta^{4}}\{\underbrace{\dot{\rho_{+}}^{2}(1-e^{2\gamma})+e^{2\gamma}}_{\Delta^{2}}\}-\frac{\ddot{\rho_{+}}}{\Delta^{4}}\{-e^{2\gamma}+\gamma^{\prime}(1-\dot{\rho_{+}}^{2})\dot{\rho_{+}}e^{2\gamma}\}
\nonumber
\end{eqnarray}
Thus,
\begin{equation}\label{rhodotdot}
\ddot{\rho_{-}}=\frac{e^{2\gamma}}{\Delta^{4}}[\ddot{\rho_{+}}+\gamma^{\prime}(1-\dot{\rho_{+}}^{2})\dot{\rho_{+}}]
\end{equation}
Now let us look at the behavior of $\Sigma$ by first calculating
its 4-velocity$(u^{\alpha})$ and then the normal vectors
$\ve{n_{\pm}}$. Clearly in $\mathcal{M}^{+}$
$u^{\alpha}=(X_{+},\dot{\rho_{+}},0,0)$ where
$X_{+}\equiv{\frac{dX_{+}}{d\tau}}$. Then from
$u^{\alpha}u_{\alpha}=-1$ we get,
\begin{equation}\label{X}
e^{2\gamma}X^{2}(1-\dot{\rho_{+}}^{2})=1 \rightarrow
X=\frac{e^{-\gamma}}{\sqrt{1-\dot{\rho_{+}}^{2}}}
\end{equation}
where we chose the plus sign for $X$. Then the normal vector
$\ve{n_{+}}$ pointing from $\mathcal{M}^{-}$ to $\mathcal{M}^{+}$
can be found by letting $n^{+}_{\alpha}=(k,l,0,0)$ and inserting
this to the equations which $\ve{n_{+}}$ must satisfy which are,
\begin{equation}\nonumber
n_{+}^{\alpha}n^{+}_{\alpha}=1, \quad n^{+}_{\alpha}u^{\alpha}=0
\end{equation}
which then gives,
\begin{equation}\label{n+}
n_{\alpha}^{+}=e^{2\gamma}X_{+}(-\dot{\rho_{+}},1,0,0)
\end{equation}
Since we are looking for surface stress-energy on $\Sigma$, we now
will calculate the \emph{Second Fundamental Forms} in
$\mathcal{M}^{-}$ and $\mathcal{M}^{-}$. Recall that,
\begin{equation}\nonumber
K_{ab}=
-n_{\alpha}(\frac{\partial^{2}{x^{\alpha}}}{\partial{\xi^{a}}\partial{\xi^{b}}}+\Gamma_{\beta\gamma}^{\alpha}\frac{\partial{x^{\beta}}}{\partial{\xi^{a}}}\frac{\partial{x^{\gamma}}}{\partial{\xi^{b}}})
\end{equation}
where $\xi^{a}=\{\tau,z,\phi\}$ are the intrinsic coordinates on
$\Sigma$. The non-vanishing components of the Christoffel symbols
in $\mathcal{M}^{-}$ is given by,
\begin{equation}
\Gamma_{\phi\phi}^{r}=-r, \quad
\Gamma_{{\phi}r}^{\phi}=\frac{1}{r}
\end{equation}
and the non-vanishing components of the Christoffel symbols in
$\mathcal{M}^{+}$ is given by,
\begin{equation}
\Gamma_{rr}^{r}=-\Gamma_{tt}^{t}=\Gamma_{tr}^{t}=-\Gamma_{rr}^{t}=\Gamma_{tt}^{r}=-\Gamma
_{tr}^{r}=\gamma^{\prime}, \quad
\Gamma_{\phi\phi}^{r}=-\frac{r}{e^{2\gamma}}, \quad
\Gamma_{r{\phi}}^{\phi}=\frac{1}{r}
\end{equation}
Thus, the non-vanishing components of $K_{ab}^{+}$ is given by,
\begin{equation}\label{Kpp+}
K_{\phi\phi}^{+}=X_{+}r, \rightarrow
K_{\hat{\phi}\hat{\phi}}^{+}\mid_{\Sigma}=\frac{X_{+}}{\rho_{+}},
\quad K_{\hat{z}\hat{z}}^{+}=0
\end{equation}
Note that we convert our tensorial quantities to their
counterparts in local Minkowski frame by
$K_{\hat{a}\hat{b}}=K_{ab}e_{\hat{a}}^{a}e_{\hat{b}}^{b}$. Now
calculation of $K_{\tau\tau}^{+}$ is a little tedious, but
straightforward. Using the definition,
\begin{eqnarray}
K_{\tau\tau}^{+} &=&
-n_{t}^{+}\{\frac{\partial^{2}{t_{+}}}{\partial{\tau^{2}}}+\Gamma_{tt}^{t}(\frac{\partial{t_{+}}}{\partial{\tau}})^{2}+\Gamma_{rr}^{t}(\frac{\partial{r}}{\partial{\tau}})^{2}+2\Gamma_{rt}^{t}(\frac{\partial{r}}{\partial{\tau}})(\frac{\partial{t_{+}}}{\partial{\tau}})\}
\nonumber \\ &-&
n_{r}^{+}\{\frac{\partial^{2}{r}}{\partial{\tau^{2}}}+\Gamma_{tt}^{r}(\frac{\partial{t_{+}}}{\partial{\tau}})^{2}+\Gamma_{rr}^{r}(\frac{\partial{r}}{\partial{\tau}})^{2}+2\Gamma_{rt}^{r}(\frac{\partial{r}}{\partial{\tau}})(\frac{\partial{t_{+}}}{\partial{\tau}})\}
\nonumber \\ &=&
e^{2\gamma}X_{+}\{\frac{dX_{+}}{d\tau}-\gamma^{\prime}X_{+}^{2}+2\gamma^{\prime}X_{+}^{2}\dot{\rho_{+}}-\gamma_{\prime}X_{+}^{2}\dot{\rho_{+}}^{2}\}
\nonumber \\ &-&
e^{2\gamma}X_{+}\{\frac{d(\dot{\rho_{+}}X_{+})}{d\tau}+\gamma^{\prime}X_{+}^{2}-2\gamma^{\prime}X_{+}^{2}\dot{\rho_{+}}+\gamma_{\prime}X_{+}^{2}\dot{\rho_{+}}^{2}\}
\nonumber \\ &=&
e^{2\gamma}X_{+}\{\underbrace{\dot{\rho_{+}}\frac{dX_{+}}{d\tau}-\frac{d(\dot{\rho_{+}}X_{+})}{d\tau}}_{-X_{+}^{2}\ddot{\rho_{+}}}+\gamma^{\prime}X_{+}^{2}[\dot{\rho_{+}}^{2}(1-\dot{\rho_{+}})-(1-\dot{\rho_{+}})]\}
\nonumber
\end{eqnarray}
where in the last line we used
$u_{\alpha}u^{\alpha}=e^{2\gamma}X_{+}^{2}(\dot{\rho_{+}}^{2}-1)=-1$.Thus
finally,
\begin{equation}\label{Ktt+}
K_{\tau\tau}^{+}=-e^{2\gamma}X_{+}^{3}\ddot{\rho_{+}}-\gamma^{\prime}X_{+}(1-\dot{\rho})
\end{equation}
We have calculated the components of the second fundamental form
in $\mathcal{M}^{+}$, let us now calculate its components in
$\mathcal{M}^{-}$. To do this we have to first find $\ve{n_{-}}$.
Clearly the 4-velocity of the shell $u^{\alpha}$ is the same in
$\mathcal{M}^{+}$ and $\mathcal{M}^{-}$, but the normal vectors
clearly change sign, since second fundamental form is a measure of
how the normal vector pointing towards the desired spacetime
changes on the hypersurface. Therefore with the same
considerations that lead us to (\ref{n+}) with the sign reversed,
we get,
\begin{equation}\label{n-}
n_{\alpha}^{-}=X_{-}(\dot{\rho_{-}},-1,0,0)
\end{equation}
(Note that $X_{-}$ has the same sign with $X_{+}$ since
$X_{-}={\Delta}X_{+}$ and $\Delta>0$) Thus the from the
Christoffel symbols we calculated for $\mathcal{M}^{-}$ above, the
non-vanishing components of the second fundamental form becomes,
\begin{equation}\label{Kpp-}
K_{\phi\phi}^{-}=-X_{-}r \rightarrow
K_{\hat{\phi}\hat{\phi}}^{-}\mid_{\Sigma}=\frac{{\Delta}X_{+}}{\rho_{+}},
\quad K_{\hat{z}\hat{z}}^{-}=0
\end{equation}
The important term $K_{\tau\tau}^{-}$ becomes,
\begin{eqnarray}
K_{\tau\tau}^{-} &=&
-n_{t}^{-}\frac{\partial^{2}{t_{-}}}{\partial{\tau^{2}}}-n_{r}^{-}\frac{\partial^{2}{r}}{\partial{\tau^{2}}}
\nonumber \\ &=&
-X_{-}(\underbrace{\dot{\rho_{-}}\frac{dX_{-}}{d\tau}-\frac{d(\dot{\rho_{-}}X_{-})}{d\tau}}_{-X_{-}^{2}\ddot{\rho_{-}}})
\nonumber
\end{eqnarray}
Therefore,
\begin{equation}\nonumber
K_{\tau\tau}^{-}=X_{-}^{3}\ddot{\rho_{-}}
\end{equation}
Now using $X_{-}={\Delta}X_{+}$ and (\ref{rhodotdot}), the above
equation becomes,
\begin{eqnarray}
K_{\tau\tau}^{-} &=&
{\Delta}^{3}X_{+}^{3}[\frac{e^{2\gamma}}{\Delta^{4}}(\ddot{\rho_{+}}+\gamma^{\prime}\dot{\rho_{+}}(1-\dot{\rho_{+}}^{2}))]
\nonumber \\ &=&
\frac{X_{+}^{3}}{\Delta}e^{2\gamma}\ddot{\rho_{+}}+\frac{\gamma^{\prime}}{\Delta}X_{+}\dot{\rho_{+}}\underbrace{e^{2\gamma}X_{+}^{2}(1-\dot{\rho_{+}}^{2})}_{=1}
\nonumber \\ &=&
\frac{X_{+}^{3}}{\Delta}e^{2\gamma}\ddot{\rho_{+}}+\frac{\gamma^{\prime}}{\Delta}X_{+}\dot{\rho_{+}}
\label{Ktt-}
\end{eqnarray}
where in the $2^{nd}$ line we used again
$u_{\alpha}u^{\alpha}=e^{2\gamma}X_{+}^{2}(\dot{\rho_{+}}^{2}-1)=-1$.
Thus in summary,
\begin{itemize}
\item[1)]
$K_{\tau\tau}^{+}=-e^{2\gamma}X_{+}^{3}\ddot{\rho_{+}}-\gamma^{\prime}X_{+}(1-\dot{\rho_{+}}),
\quad
K_{\tau\tau}^{-}=\frac{X_{+}^{3}}{\Delta}e^{2\gamma}\ddot{\rho_{+}}+\frac{\gamma^{\prime}}{\Delta}X_{+}\dot{\rho_{+}}$
\item[2)] $
K_{\hat{\phi}\hat{\phi}}^{+}\mid_{\Sigma}=\frac{X_{+}}{\rho_{+}},
\quad
K_{\hat{\phi}\hat{\phi}}^{-}\mid_{\Sigma}=-\frac{{\Delta}X_{+}}{\rho_{+}}$
\item[3)] $K_{\hat{z}\hat{z}}^{-}=0, \quad
K_{\hat{z}\hat{z}}^{+}=0$
\end{itemize}
since we have now finished the calculation of second fundamental
forms, now we can use the famous formula connecting $[K_{ab}]$ to
the surface stress-energy $S_{ab}$ where $[{\quad}]$ represents
for an quantity $\Psi$, we have $[\Psi]=\Psi^{+}-\Psi^{-}$.
\begin{equation}\label{Sij}
S_{ab}=-\frac{1}{8\pi}\{[K_{ab}]-g_{ab}[K]\}
\end{equation}
Therefore, from the second fundamental forms we obtained above, we
see that the surface stress-energy tensor is diagonal with its
elements
$S_{\hat{a}\hat{b}}=diag[S_{\tau\tau},S_{\hat{\phi}\hat{\phi}},S_{\hat{z}\hat{z}}]$.
Calculation of these elements are trivial. The traces of the
second fundamental forms and the discontinuity is given by,
\begin{eqnarray}
K^{+} &=& -K_{\tau\tau}^{+}+K_{\hat{\phi}\hat{\phi}}^{+} \nonumber \\
K^{-} &=& -K_{\tau\tau}^{-}+K_{\hat{\phi}\hat{\phi}}^{-} \nonumber \\
{[K]} &=& -[K_{\tau\tau}]+[K_{\phi\phi}] \label{K}
\end{eqnarray}
Then using (\ref{K}) in (\ref{Sij}), we have the following
equations,
\begin{eqnarray}
S_{\tau\tau} &=& -\frac{[K_{\hat{\phi}\hat{\phi}}]}{8\pi} \label{Stt}  \\
S_{\hat{\phi}\hat{\phi}} &=& -\frac{[K_{\tau\tau}]}{8\pi}
\label{Spp} \\ S_{\hat{z}\hat{z}} &=&
-\frac{1}{8\pi}\{[K_{\tau\tau}]-[K_{\hat{\phi}\hat{\phi}}]\}
\label{Szz}
\end{eqnarray}
Thus in full from by letting $S_{\tau\tau}=\epsilon, \
S_{\hat{\phi}\hat{\phi}}=p_{\phi}, \ S_{\hat{z}\hat{z}}=p_{z}$, we
have,
\begin{itemize}
\item[1)]
$\epsilon=-\frac{1}{8\pi}\frac{X_{+}}{\rho_{+}}(\Delta+1)$
\item[2)]
$p_{\phi}=\frac{1}{8\pi}\{e^{2\gamma}X_{+}^{3}\ddot{\rho_{+}}(1+\frac{1}{\Delta})+\gamma^{\prime}X_{+}\}$
\item[3)]
$p_{z}=\frac{1}{8\pi}\{e^{2\gamma}X_{+}^{3}\ddot{\rho_{+}}(1+\frac{1}{\Delta})+\gamma^{\prime}X_{+}+\frac{X_{+}}{\dot{\rho_{+}}}(\Delta+1)\}$
\end{itemize}
Instead of following the route of ~\cite{MS} and ~\cite{PW}, in
which the dynamical behavior of the above equations are
investigated to see how $\ddot{\rho_{+}}$ changes to comment on
whether the shell collapses or not, we will consider conservation
equations of Israel~\cite{I} which were derived for vacuum and in
~\cite{MK} for arbitrary spacetimes. The conservation equations
are,
\begin{itemize}
\item[1)] $\widetilde{K}^{b}_{a|b}-\widetilde{K}_{|a}=
8\pi{\widetilde{(T_{\alpha\beta}e_{a}^{\alpha}n^{\beta})}}$
\item[2)] $S_{a|b}^{b} =
-{[T_{\alpha\beta}e_{a}^{\alpha}n^{\beta}]}$ \item[3)]
$S^{ab}\widetilde{K}_{ab} = [T_{\alpha\beta}n^{\alpha}n^{\beta}]$
\item[4)]
${}^{(3)}R+\widetilde{K}_{ab}\widetilde{K}^{ab}-\widetilde{K}^{2}
= -\frac{1}{4}(8\pi)^{2}(S_{ab}S^{ab}-\frac{S^{2}}{2}) -
8\pi{\widetilde{(T_{\alpha\beta}n^{\alpha}n^{\beta})}}$
\end{itemize}
Let us consider the $3^{rd}$ conservation equation which will
prove an incredible result. Since the right hand side of this
equation contains the discontinuity of the stress-energy tensor on
$\mathcal{M}^{+}$ and $\mathcal{M}^{-}$ we have to calculate it
from the Einstein Field Equations
$G^{\pm}_{\alpha\beta}=8{\pi}T^{\pm}_{\alpha\beta}$. The interior
spacetime $\mathcal{M}^{-}$ is flat therefore making
$T^{-}_{\alpha\beta}=0$. On the other hand we have,
\begin{equation}\label{einstein}
G_{\alpha\beta}^{+}=\frac{\gamma^{\prime}}{r}\zeta_{\alpha}\zeta_{\beta}
\end{equation}
where $\zeta_{\alpha}=(1,-1,0,0)$ is a null vector. This clearly
represents a spacetime with radiation of null-particles.
Therefore, the stress-energy tensors become,
\begin{eqnarray}
T^{+}_{\alpha\beta} &=&
\frac{\gamma^{\prime}}{8{\pi}r}\zeta_{\alpha}\zeta_{\beta}
\label{T+} \\ T^{-}_{\alpha\beta} &=& 0 \label{T-}
\end{eqnarray}
Thus, given (\ref{n+}) and (\ref{T+}) we get,
\begin{equation}\label{Tn}
[T_{\alpha\beta}n^{\alpha}n^{\beta}]=\frac{\gamma^{\prime}}{8{\pi}\rho_{+}}X_{+}^{2}(\dot{\rho_{+}}-1)^{2}
\end{equation}
The left hand side of the $3^{rd}$ conservation equations can be
calculated by just inserting the equations
(\ref{Ktt+})-(\ref{Ktt-})-(\ref{Kpp+})-(\ref{Kpp-}) and
(\ref{Stt})-(\ref{Spp}) just derived above yielding,
\begin{eqnarray}
\widetilde{K}_{\hat{a}\hat{b}}S^{\hat{a}\hat{b}} &=&
\widetilde{K}_{\tau\tau}S^{\tau\tau}+\widetilde{K}_{\hat{\phi}\hat{\phi}}S^{\hat{\phi}\hat{\phi}}
\nonumber \\ &=&
\frac{1}{16\pi}(K^{+}_{\tau\tau}+K^{-}_{\tau\tau})(K^{-}_{\hat{\phi}\hat{\phi}}-K^{+}_{\hat{\phi}\hat{\phi}})+\frac{1}{16\pi}(K^{+}_{\hat{\phi}\hat{\phi}}+K^{-}_{\hat{\phi}\hat{\phi}})(K^{-}_{\tau\tau}-K^{+}_{\tau\tau})
\nonumber
\end{eqnarray}
Thus we have,
\begin{equation}\label{cons}
-K^{+}_{\tau\tau}K^{+}_{\hat{\phi}\hat{\phi}}+K^{-}_{\tau\tau}K^{-}_{\hat{\phi}\hat{\phi}}=\frac{\gamma^{\prime}X_{+}^{2}(\dot{\rho_{+}}-1)^{2}}{\rho_{+}}
\end{equation}
Thus, inserting
(\ref{Ktt+})-(\ref{Ktt-})-(\ref{Kpp+})-(\ref{Kpp-}) in the above
equation we get,
\begin{equation}\nonumber
\{e^{2\gamma}X_{+}^{3}\ddot{\rho_{+}}+\gamma^{\prime}X_{+}(1-\dot{\rho_{+}})\}(\frac{X_{+}}{\rho_{+}})-(\frac{{\Delta}X_{+}}{\rho_{+}})\frac{1}{\Delta}\{e^{2\gamma}X_{+}^{3}\ddot{\rho_{+}}+\gamma^{\prime}X_{+}\dot{\rho_{+}}\}=\frac{\gamma^{\prime}X_{+}^{2}(\dot{\rho_{+}}-1)^{2}}{\rho_{+}}
\end{equation}
Then we get the trivial solution for $\dot{\rho_{+}}$ (note that
$\gamma^{\prime}{\ne}0$ is assumed, otherwise both metrics would
be flat),
\begin{equation}\label{resultrho}
\dot{\rho_{+}}=0
\end{equation}
Thus the shell never collapses and just stays at constant radius.
This means that there can not be a shell motion, which is contrary
to the results of~\cite{MS}-~\cite{PW}.

The results we obtained above can also be obtained by the use of
Hamiltonian formalism as done by Olea and Cris\'{o}stomo
~\cite{CO} for the spherically symmetric case. The variation of
the gravitational action yields the Hamiltonian and the momentum
constraints which can be calculated for both $\mathcal{M}^{+}$ and
$\mathcal{M}^{-}$. Then by integrating these constraints across
$r=\rho_{+}$ one obtains the same equations that have been
obtained above by the conservation equations so both approaches
yield the same results. (Actually the conservation equations of
Israel are obtained by adding and subtracting the Hamiltonian and
momentum constraints corresponding to $\mathcal{M}^{+}$ and
$\mathcal{M}^{-}$. The distinction is that in Israel formalism,
these constraints are obtained by the Gauss-Codazzi Equations
rather than the variational principle).

Now let us look at the $2^{nd}$ conservation equation. First of
all, the right hand side is given by using
$u^{\alpha}=(X_{+},\dot{\rho_{+}},0,0)$ and equation (\ref{n+})
\begin{eqnarray}
T_{\alpha\beta}^{+}e_{a}^{\alpha}n_{+}^{\beta} &=&
\frac{\gamma^{\prime}}{8{\pi}r}(\zeta_{\beta}n^{\beta})(\zeta_{\alpha}u^{\alpha})=-\frac{\gamma^{\prime}X_{+}^{2}}{8{\pi}\rho_{+}}
\quad \textrm{for} \ a=\tau \nonumber \\ &=& 0 \quad \textrm{for}
\ a=\phi,z \nonumber
\end{eqnarray}
since
$e_{\phi}^{\alpha}\zeta_{\alpha}=e_{z}^{\alpha}\zeta_{\alpha}=0$
and $\dot{\rho_{+}}=0$. Then the nontrivial part of $2^{nd}$
conservation equation becomes,
\begin{equation}\nonumber
S_{\tau|b}^{b}=S_{\tau|\tau}^{\tau}=\frac{\gamma^{\prime}X_{+}^{2}}{8{\pi}\rho_{+}}
\end{equation}
Using (\ref{Stt}) we have
\begin{equation}\nonumber
S_{\tau}^{\tau}=\frac{[K_{\hat{\phi}\hat{\phi}}]}{8\pi}=\frac{X_{+}(1+\Delta)}{8{\pi}\rho_{+}}
\end{equation}
Note that from (\ref{delta}) when we have for $\dot{\rho_{+}}=0$
we get $\Delta=e^{\gamma}$ and from (\ref{X}), we have
$X_{+}=e^{-\gamma}$.Therefore, noting that the intrinsic covariant
derivative becomes just the ordinary derivative for
$\tau$-component from (\ref{induced}), we get,
\begin{equation}\nonumber
S_{\tau|\tau}^{\tau}=\{\frac{e^{-\gamma}+1}{8{\pi}\rho_{+}}\}_{|\tau}=\frac{\gamma^{\prime}e^{-2\gamma}}{8{\pi}\rho_{+}}
\end{equation}
Where we used the fact that
$\frac{d\gamma}{d\tau}=X_{+}\dot{\gamma}=-e^{-\gamma}\gamma^{\prime}$
since we had $\gamma^{\prime}=-\dot{\gamma}$. Thus finally we get
\begin{equation}\label{trivial}
S_{\tau|\tau}^{\tau}=\frac{\gamma^{\prime}e^{-2\gamma}}{8{\pi}\rho_{+}}=-{[T_{\alpha\beta}e_{a}^{\alpha}n^{\beta}]}
\end{equation}
by which we get trivially $1=1$ which justifies our calculations.

Now we have not used the $1^{st}$ and $4^{th}$ conservation
equations, since the $1^{st}$ one when applied results in equation
(\ref{trivial}) and the $4^{th}$ one results in equation
(\ref{resultrho}).

Clearly, we can not go to the null limit as was done in ~\cite{MS}
since we have found that the shell is stationary. Therefore, we
can state that the null case must be directly considered by the
null-shell formalism separately as we did, rather than looking at
the null limit of the timelike case.

Here we would like to remark that we are at a variance of the
results in ~\cite{MS}-\cite{PW}. For example in ~\cite{MS}, it was
found that both in the timelike case and in the lightlike limit,
the shell first approaches to a minimum radius and then expands
infinitely. However, we found out that in the timelike case the
shell is stationary and in the lightlike limit we get a collapsing
shell. In addition, if we impose cylindrical flatness, for the
null case we get a boundary layer so the shell is lost.

The difference of our work and previous ones is due to the fact
that they did not considered the conservation equations at all and
some difference in the calculated values of the second fundamental
forms. We can compare our results with the results of ~\cite{MS}
now since it is the most recent work. The extrinsic curvatures for
the spacetimes (\ref{-manif}) and (\ref{+manif}) joined at
$\Sigma$ was found as,
\begin{eqnarray}
K_{\tau\tau}^{+} &=&
-X_{+}^{3}e^{2\gamma}\ddot{\rho_{+}}+\gamma^{\prime}X_{+}^{3}\dot{\rho_{+}}^{2}(1-\dot{\rho_{+}})
\label{mist1} \\ K_{\tau\tau}^{-} &=&
-X_{-}^{3}\ddot{\rho_{-}}=-\frac{X_{+}^{3}}{\Delta}e^{2\gamma}\ddot{\rho_{+}}-\frac{\gamma^{\prime}}{\Delta}X_{+}^{3}e^{2\gamma}(1-\dot{\rho_{+}}^{2})\dot{\rho_{+}}
\label{mist2} \\ K_{\tau\tau}^{\pm} &=& 0 \nonumber \\
K^{+}_{\hat{\phi}\hat{\phi}} &=& \frac{X_{+}}{\rho_{+}} \label{mist3} \\
K_{\hat{\phi}\hat{\phi}}^{-} &=& \frac{{\Delta}X_{+}}{\rho_{+}}
\label{mist4}
\end{eqnarray}
Using the fact that
$X_{+}^{2}e^{2\gamma}(1-\dot{\rho_{+}}^{2})=1$, equation
(\ref{mist2}) reduces to $K_{\tau\tau}^{-}=
-\frac{X_{+}^{3}}{\Delta}e^{2\gamma}\ddot{\rho_{+}}-\frac{\gamma^{\prime}}{\Delta}X_{+}\dot{\rho_{+}}$.
First of all, one can see the most apparent difference with our
results with Seriu's ~\cite{MS} is that the sign of the normal
vector $\ve{n}{\perp}\Sigma$ is taken to be $(+)$ in both
$\mathcal{M}^{+}$ and in $\mathcal{M}^{-}$ (note that
$X_{+},X_{-}>0$) which the correct selections should be as ours
(i.e negative sign in $\mathcal{M}^{-}$ and positive sign in
$\mathcal{M}^{+}$). Secondly, the equation for $K_{\tau\tau}^{+}$
is different as can be compared with our previous calculations.
Let us show that these results lead to a contradiction in the
conservation equations. Putting these values in (\ref{cons}) we
get,
\begin{equation}\nonumber
[e^{2\gamma}X_{+}^{3}\ddot{\rho_{+}}-\gamma^{\prime}X_{+}^{3}e^{2\gamma}\dot{\rho_{+}}^{2}(1-\dot{\rho_{+}})](\frac{X_{+}}{\rho_{+}})-(\frac{{\Delta}X_{+}}{\rho_{+}})\frac{1}{\Delta}[e^{2\gamma}X_{+}^{3}\ddot{\rho_{+}}+\gamma^{\prime}X_{+}\dot{\rho_{+}}]=\frac{\gamma^{\prime}X_{+}^{2}(\dot{\rho_{+}-1})^{2}}{\rho_{+}}
\end{equation}
which reduces to
\begin{equation}\label{mist5}
-(X_{+}^{2}e^{2\gamma}\dot{\rho_{+}}^{2}(1-\dot{\rho_{+}})+\dot{\rho_{+}})=\frac{(\dot{\rho_{+}}-1)^{2}}{\rho_{+}}
\end{equation}
Now using $X_{+}^{2}e^{2\gamma}(1-\dot{\rho_{+}}^{2})=1$ once more
in the above equation we get,
\begin{equation}\label{fin}
\frac{\dot{\rho_{+}}^{2}}{1+\dot{\rho_{+}}}+\dot{\rho_{+}}=-\frac{(\dot{\rho_{+}}-1)^{2}}{\rho_{+}}
\end{equation}
In principle, the above expression can be rewritten by solving the
cubic equation for $\dot{\rho_{+}}$, then this equation can be
integrated. Doing so will prove to be very difficult, since the
integral to be evaluated is very complicated. Rather than doing
this, we can analyze the equation (\ref{fin}) without solving it.
Clearly, we must have ${\mid}{\dot{\rho_{+}}}{\mid}<1$ since the
shell is assumed to be timelike. For the expansion of the shell,
$0<\dot{\rho_{+}}<1$, there is no solution, since the right hand
side is always negative, but the left hand side is always positive
which is contradictory. For the collapsing shell
$-1<\dot{\rho_{+}}<0$, the contradiction can only be removed if
$-1<\dot{\rho_{+}}<-0.5$ is imposed. However, this result is a
contradiction since the only possible solution Seriu ~\cite{MS}
obtains is the expanding shell, where $\dot{\rho_{+}}>0$.

\section{Conclusion}

In this work we considered an axial-symmetric hypersurface
separating an interior flat and an exterior curved radiation
filled spacetimes both in the null and the timelike cases. We
found that in the null case the shell collapses to zero radius at
the speed of light but when cylindrical flatness is imposed the
shell is lost (i.e. surface stress-energy becomes identically
zero). and a boundary layer is obtained. For the timelike case, we
used the full conservation equations to determine the dynamical
behavior of a thin shell for the first time, and showed that the
axial-symmetric shell in consideration stays stationary at some
constant radius. The Hamiltonian formalism presented in ~\cite{CO}
for spherically symmetric shells, also gives the same results when
used for the configuration we analyze.

With these results, we are at a variance with the predictions of
~\cite{PW} and ~\cite{MS} since we considered the conservation
equations to determine the dynamical behavior rather than the
equations on stress-energy. Seriu ~\cite{MS} found that in the
timelike case and its null limit, the shell first contracts to a
minimum nonzero radius and then expands infinitely. Pereira and
Wang ~\cite{PW} also obtained similar, but they also found
collapsing shell configurations in some special cases. Both of
these results are different from ours for the reasons we discussed
above. However, one should consider the conservation equations
when analyzing the dynamical behavior of thin shells since they
are powerful constraints on the equations that are obtained from
the surface stress-energy.

\section{Acknowledgments}

It is a pleasure to express my indebtedness to Professor Metin
G\"{u}rses who stimulated my interest in axial-symmetric shells.

\end{document}